\documentstyle[twoside,fleqn,espcrc2,epsf]{article}

\newcommand{\AmS}{{\protect\the\textfont2
  A\kern-.1667em\lower.5ex\hbox{M}\kern-.125emS}}

\title{Large scale numerical simulation of the three-state Potts model}

\author{Shigemi Ohta
	\address{The Institute of Physical and Chemical Research (RIKEN),
	         Wako-shi, Saitama 351-01, Japan}%
        \thanks{This research is supported in part by the Parallel
		Computing Research Facility, Fujitsu Laboratory, who
		provided the computing time on their experimental
		parallel computer, AP1000.}}

\begin{document}

\begin{abstract}
The three-state Potts model is numerically investigated on
three-dimensional simple cubic lattices of up to \(128^3\) volume,
concentrating on the neighborhood of the first-order phase transition
separating the ordered and disordered phases.  The ordered phase is
found to allow admixture of disordered domains induced by a
long-range attraction acting between the two different non-favored
spins.  This phenomenon gives an explanation of why the first-order
phase transitions associated with the global \(Z_3\) symmetry are so
weak.
\end{abstract}

\maketitle

\section{INTRODUCTION}

It is by now established that the systems with global \(Z_3\) symmetry
in three spatial dimensions at finite temperature have in common a
weak first-order phase transition separating ordered and disordered
phases.

Lattice numerical calculations of the pure-gauge \(SU(3)\) quantum
chromodynamics (QCD) at finite temperature \cite{ref:puregauge}
revealed a first-order phase transition that separates a
low-temperature color-confining phase and a high-temperature
non-confining phase.  The confining phase is disordered with the
Polyakov line order parameter taking zero expectation value.  The
non-confining phase is ordered and the Polyakov line takes a finite
expectation value the argument of which falls on one of the three
\(Z_3\) axes in the complex plane.  The transition is considered weak
because the latent heat is small and the Polyakov line correlation
length grows.

Numerical simulations of the three-state Potts model \cite{ref:Potts}
showed that there is a first-order phase transition separating a
low-temperature ordered phase and a high-temperature disordered one.
This transition is weak in the sense the latent heat is small.

However, it is not yet understood why these phase transitions are so
weak.  The author investigated this problem through numerical
simulations of the three-state Potts model on simple cubic lattices
with up to \(128^3\) volumes.

\section{POTTS MODEL}

We consider the simplest form of the three-state Potts model on
three-dimensional simple cubic lattices.  At each site, \(x\), of the
lattice, a spin, \(\sigma_x\), is defined. It takes one of the three
possible states in the \(Z_3\) group, \(\sigma_x \in \{0, 1, 2\}\).
The spins interact through a nearest-neighbor interaction described by
the Hamiltonian \begin{equation} H = -K \sum_{\langle xy \rangle}
\delta(\sigma_x, \sigma_y). \end{equation} The model is invariant
under global \(Z_3\) transformations.

We are only interested in ``ferromagnetic'' couplings, \(K > 0\),
because the effective interactions among the Polyakov lines in the
finite-temperature pure-gauge QCD are of this type \cite{ref:FOU}.

Thermodynamics of the model is described by the partition function
\begin{equation} Z = \exp(-H/T) = \exp(-J \sum_{\langle xy \rangle}
\delta(\sigma_x, \sigma_y)), \end{equation}  where the dimensionless
coupling is defined as \(J = K/T\).

\section{NUMERICAL SIMULATION}

Earlier numerical simulations showed that the system undergoes a weak
first-order phase transition that separates a low-temperature ordered
phase and a high-temperature disordered phase at \(J \simeq 0.5505\).
It is considered weak because the dimensionless internal energy
density \begin{equation} \epsilon = \sum_{\langle xy \rangle}
\delta(\sigma_x, \sigma_y) / \sum_{\langle xy \rangle} 1
\end{equation} shows only a small gap at the phase transition, from
about 0.58 to 0.53, compared with the maximum possible gap from 1 to
1/3.

However, these results were obtained on relatively small volumes such
as \(32^3\).  On such small volumes flip-flop transitions between the
two phases occur frequently, and it is hard to tell in which of the
two phases the system resides at a given instant.  Because of this
difficulty, the natures of the individual phases could not be studied,
nor the reason why the phase transition is so weak.

To avoid this difficulty one has to simulate the system on larger
volumes.  The author wrote a heatbath simulation code for the model
that runs on an experimental parallel computer, AP1000, built by
Fujitsu Laboratory \cite{ref:AP1000}.  On the largest existing
configuration of the computer with \(16 \times 32 = 512\)
microcomputer ``cells,'' the code performs one million heatbath
updates of a \(128^3\) volume in about 32 hours.

Typically a couple of million heatbath updates are made at four values
of the coupling, \(J = 0.55025\), 0.5505, 0.55075, and 0.551.  At each
of these coupling values, two independent simulations were made; one
starting from a completely ordered configuration and the other
starting from a completely disordered one.

At the highest temperature, \(J = 0.55025\), the ordered-start
simulation quickly converged to the disordered phase within 100K
heatbath updates, while the disordered-start simulation remained in
the disordered phase indefinitely through the duration of the
simulation for more than one million updates.  In contrast, at the
lowest temperature, \(J = 0.551\), the disordered-start simulation
quickly converged to the ordered phase while the ordered-start one
remained in the ordered phase indefinitely.  Thus we narrowed down the
phase transition temperature, \(J_c\), between these two values,
\(0.55025 < J_c < 0.551\).

Coexistence of the two phases is observed at both of the intermediate
temperatures, \(J = 0.5505\) and 0.55075: The ordered-start
simulations remained in the ordered phase and the disordered-start
ones remained in the disordered phase through their durations of more
than two million heatbath updates.  In each of these simulations there
is no doubt about in which of the phases the system resides: We are
now able to investigate the natures of the individual phases.

\section{CORRELATION FUNCTION}

The two-point correlation function, \(C_{ij} (r)\), is defined as the
probability to find a pair of spin \(i\) and \(j\) separated by the
distance \(r\).

In the disordered phase the correlations decay exponentially in Yukawa
form \begin{equation} C_{ij} (r) \rightarrow \alpha_{ij}
\frac{\exp(-m_{ij}r)}{r} + p_i p_j, \end{equation} as the distance
\(r\) increases toward infinity.  The constant term is given by the
product of probabilities, \(p_i\) and \(p_j\), to find a site with
spin \(i\) and \(j\) respectively (in the disordered phase they should
all be 1/3.)  We numerically confirmed that all the correlations are
fitted well by the Yukawa-form (Table 1.)  The positive diagonal
amplitudes, \(\alpha_{ii} > 0\), shows the like spins attract each
other, in accordance with the ferromagnetic interaction.  The
corresponding correlation mass of \(m_{ii} \simeq 0.15\) suggests
there would be clusters of like spins of the size of several lattice
spacings.  Indeed in the spin distribution one sees many such
clusters.

\begin{table}
\setlength{\tabcolsep}{1.5pc}
\newlength{\digitwidth} \settowidth{\digitwidth}{\rm 0}
\catcode`?=\active \def?{\kern\digitwidth}
\caption{Two-point correlations in the disordered phase at \protect\(J
= 0.55050\protect\).  Two-parameter least-\protect\(\chi^2\protect\)
fit to the Yukawa form.}
\label{tab:J55050d}
\begin{tabular}{lrr}
\hline
\protect\((ij)\protect\)
& \multicolumn{1}{l}{\protect\(m\protect\)}
& \multicolumn{1}{l}{\protect\(\alpha\protect\)} \\
\hline
(00) & 0.15(1) & 0.071(3) \\
(11) & 0.15(1) & 0.072(2) \\
(22) & 0.15(1) & 0.071(2) \\
(01) & 0.16(1) & -0.037(1) \\
(02) & 0.16(1) & -0.037(1) \\
(12) & 0.16(1) & -0.037(1) \\
\hline
\end{tabular}
\end{table}

The behavior of the correlations in the ordered phase is not known.
The current numerical results show the Yukawa form fits them well
also in this case (Table 2.)

\begin{table}
\setlength{\tabcolsep}{1.5pc}
\caption{Two-point correlations in the ordered phase at \protect\(J
= 0.55050\protect\).  The favored spin is 0.}
\label{tab:J55050o}
\begin{tabular}{lrr}
\hline
\protect\((ij)\protect\)
& \multicolumn{1}{l}{\protect\(m\protect\)}
& \multicolumn{1}{l}{\protect\(\alpha\protect\)} \\
\hline
(00) & 0.14(2) & 0.081(3) \\
(11) & 0.18(3) & 0.028(4) \\
(22) & 0.18(3) & 0.028(4) \\
(01) & 0.15(1) & -0.041(1) \\
(02) & 0.15(1) & -0.041(1) \\
(12) & 0.03(3) & 0.010(1) \\
\hline
\end{tabular}
\end{table}

All the diagonal correlations (\(i = j\)) are fitted well (Figure 1.)
The positive amplitudes, \(\alpha_{ii} > 0\), are in accordance with
the ferromagnetic interaction.  Smaller correlation mass for the
favored spin, 0, compared with those of non-favored spins, 1 and 2, is
consistent with the fact that the favored spin dominates the volume in
the ordered phase.

\begin{figure}
\leavevmode
\epsfxsize=75mm
\epsffile{fig1.eps}
\caption{Diagonal correlations in the ordered phase at \protect\(J =
0.5505\protect\).  The constant terms are subtracted.  The fitting
curves are from Table 2.}
\label{fig:J5505olog}
\end{figure}

The correlations between the favored spin and either of the two
non-favored spins are also fitted well by the Yukawa form (Figure
2.)  Now the negative amplitudes, \(\alpha_{01} < 0\) and
\(\alpha_{02} < 0\), means the domains of favored spin repel the
non-favored spins.

\begin{figure}
\leavevmode
\epsfxsize=75mm
\epsffile{fig2.eps}
\caption{Offdiagonal correlations in the ordered phase at \protect\(J
= 0.5505\protect\).}
\label{fig:J5505ooff}
\end{figure}

The correlation between the two different non-favored spins, \(C_{12}
(r)\), is the most interesting.  It starts from zero at the origin as
it should be, but then overshoots the asymptotic value \(p_1 p_2\),
and approaches it from above (Figure 2.)  This of course means a
long-range attractive force acts between the two different non-favored
spins.  This long-range attractive part alone can be fitted by the
Yukawa form(Table 2.)  Note that the correlation mass, \(m_{12} = 0.03
\pm 0.03\), is much smaller than those for the other cases, the
smallest of which is \(m_{00} = 0.14 \pm 0.02\).

This long-range attraction between the two different non-favored spins
is likely to cause instability of the ordered phase dominated by the
favored spin.  Indeed in the spin distribution of this phase one sees
that there are many clusters of the non-favored spins, and that a
cluster of a non-favored spin almost always accompany clusters of the
other non-favored spin in its neighborhood.  See Figure 3 for more
detail, in which three cross sections of a typical spin distribution
on the \(128^3\) volume is shown with the favored spin in white and
the non-favored ones in grey and black.  It should be noted that such
pairing of the clusters of the two different non-favored spin is an
economical way to maintain the global \(Z_3\) symmetry which dictates
the volumes occupied by the two spins must be equal.

\begin{figure}
\leavevmode
\epsfxsize=75mm
\epsffile{fig3.eps}
\caption{Typical spin distribution in the ordered phase at \protect\(J
= 0.5505\protect\).}
\label{fig:J5505o_100}
\end{figure}

Furthermore, these clusters of the non-favored spins do not have
smooth convex shape, but have complex concave boundaries (Figure 3.)
Thus the neighborhood of such a cluster of clusters of the non-favored
spins can be considered as an island of the disordered phase in the
sea of the ordered phase.  Near the phase transition, there is so much
of such admixture of the domains of the disordered phase as the
entropy density of the ordered phase does not differ much from that of
the disordered phase, giving a natural explanation why the latent heat
is so small.

Irregular and concave boundaries of the non-favored clusters and
disordered domains in the ordered phase near the transition may be
consistent with very small surface tension at the confined-deconfined
phase boundaries found by a phenomenological MIT-bag calculation
\cite{ref:Mardor} as well as by lattice QCD numerical simulations
\cite{ref:BandH}.  But the irregular boundaries themselves suggests
inadequacy of the sharp spherical boundaries assumed or imposed in
these calculations.

The long-range correlation between the two different non-favored spins
remains attractive down to the lowest temperature simulated, \(J =
0.551\).  Its range, however, decreases to \(m_{12} = 0.07(3)\) at \(J
= 0.55075\) and 0.11(3) at 0.551.  This way the ordered phase away
from the first-order transition consolidates itself.

\section{SUMMARY}

The three-state Potts model in three spatial dimensions is numerically
investigated in the neighborhood of its first-order phase transition.
The large lattice volumes of up to \(128^3\) gave a good phase
separation.  The natures of the individual ordered and disordered
phases are studied for the first time.  In the ordered phase a long
range attraction between the two different non-favored spins is found,
probably as a direct consequence of the global \(Z_3\) symmetry
\cite{ref:fullpaper}.  Near the phase transition this attraction
induces instability of the ordered phase allowing admixture of domains
of the disordered phase.  This phenomenon gives an explanation of why
the first-order phase transitions associated with the global \(Z_3\)
symmetry are so weak.

\end{document}